\documentclass[aps,pra,showpacs,twocolumn,superscriptaddress]{revtex4-2}
\usepackage{amsfonts}
\usepackage{amssymb}
\usepackage{amsmath}
\usepackage{graphicx}
\usepackage{epsfig}
\usepackage{color}
\usepackage{amsmath,bm}
\usepackage{booktabs}
\usepackage[colorlinks=true, linkcolor=blue, citecolor=blue, urlcolor=blue]{hyperref}

\begin{document}

\title{Condensate states in Fermi and Bose-Hubbard ladders}
\affiliation{School of Physics, Nankai University, Tianjin 300071, China} %
\affiliation{School of Physics, Tianjin University, Tianjin 300072, China}
\author{F. X. Liu}
\affiliation{School of Physics, Tianjin University, Tianjin 300072, China}
\author{E. S. Ma}
\affiliation{School of Physics, Nankai University, Tianjin 300071, China}
\author{Z. Song}
\email{songtc@nankai.edu.cn}
\affiliation{School of Physics, Nankai University, Tianjin 300071, China}

\begin{abstract}
Although neither hardcore bosons nor fermions can occupy the same
single-site state, they still obey different statistics, resulting in
distinct many-particle quantum states, such as condensate states versus
Fermi-liquid states. However, when only pair states are considered, the two
can take the same form, since a local hardcore Bose pair and a Fermi pair
obey the same statistics. In this work we demonstrate this by studying both
Fermi and Bose extended Hubbard ladders, which can be realized
experimentally in synthetic atomic ladders. A set of exact condensate-pair
eigenstates for the Fermi ladder is constructed under SU(2) symmetry and can
then be obtained by the spectrum generating algebra. The corresponding
hardcore boson counterpart can be simply obtained by replacing fermionic
operators with hardcore bosonic ones. Nevertheless, the boson-pair
eigenstates are associated not with symmetry but with the restricted
spectrum generating algebra. We also investigate the effect of
next-nearest-neighbor hopping on the condensate states through numerical
simulations of the dynamic response. The conclusions can be extended to a
two-layer system. Our result reveals not only the resemblance of fermions to
hardcore bosons, but also a possible mechanism of Hilbert-space
fragmentation.
\end{abstract}

\maketitle

\section{Introduction}

Bosons and fermions are the two fundamental classes of particles in quantum
physics, distinguished by their spin and behavior. Bosons can occupy the
same quantum state, while two identical fermions cannot exist in the same
state. Apparently, bosons are social particles and fermions are solitary
particles. A direct demonstration is the Bose-Einstein condensation (BEC) 
\cite{bose1924plancks,einstein1924quantentheorie}, which serves as a
striking example of quantum phenomena that become evident on a macroscopic
scale. Specifically, BEC is marked by the formation of a coherent quantum
state among a collection of free bosons, resulting in a remarkable
synchronization of their behavior. This phenomenon never appears a
collection of free fermions. Nevertheless, when the interaction is
considered, it becomes a little complicated. On the one hand, neither
hardcore bosons nor fermions can occupy the same single-site state. On the
other hand, the Bardeen--Cooper--Schrieffer (BCS) theory describes
superconductivity as a microscopic effect caused by a condensation of Cooper
pairs \cite{Cooper1956,BCS1957a,BCS1957b}.

\begin{figure}[!ht]
\centering
\includegraphics[width=0.48\textwidth]{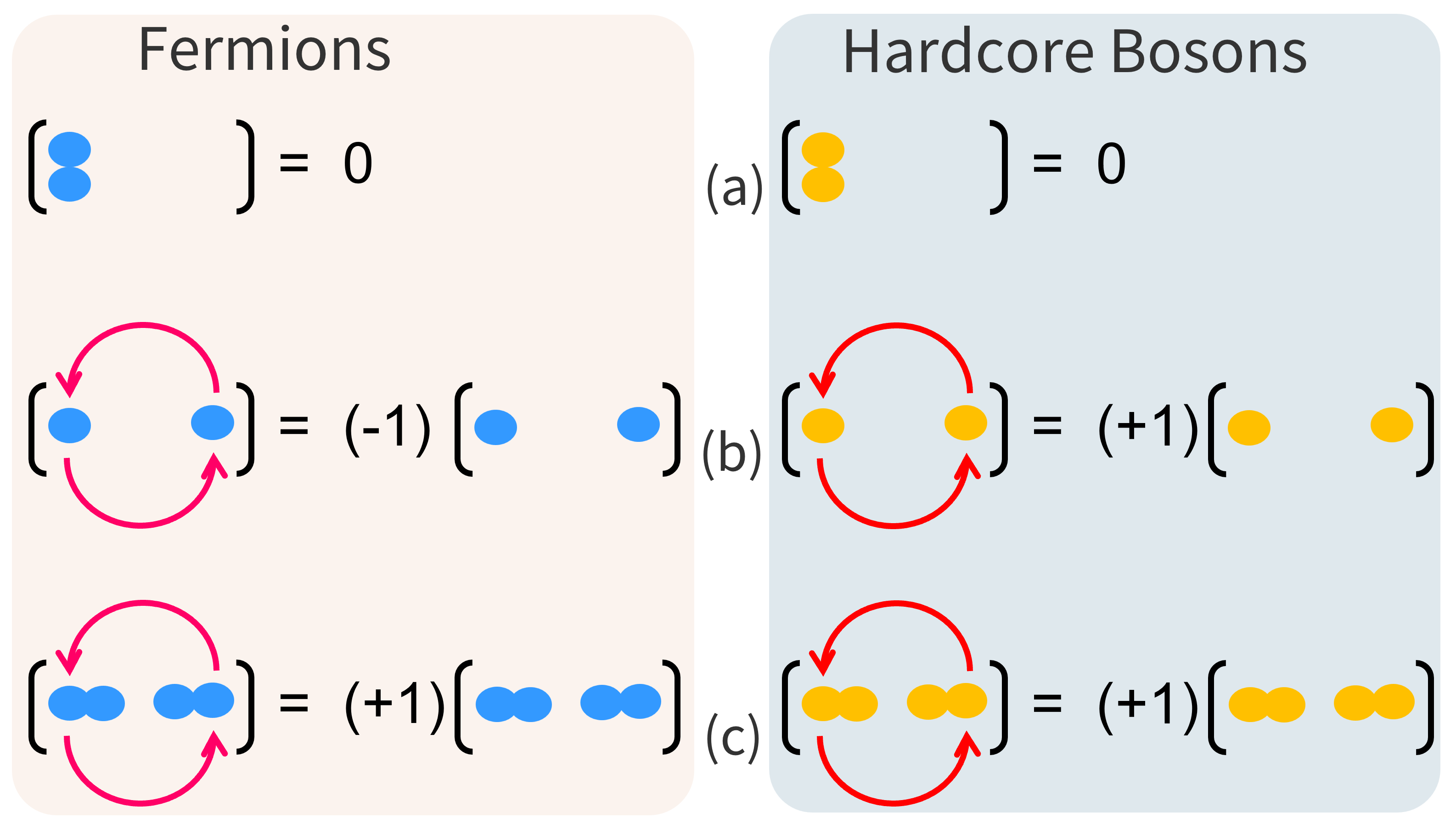}
\caption{Comparison of the statistics of fermions and hardcore bosons. (a)
Neither hardcore bosons nor fermions can occupy the same single-site state.
(b) They obey different statistics when two single particles exchange their
positions. (c) However, when only pair states are considered, a local
hardcore Bose pair and a Fermi pair obey the same statistics. This suggests
that the hardcore boson counterpart of a fermionic model may possess pair
eigenstates of the same form.}
\label{fig1}
\end{figure}

A paradigram example for singlet pair is the $\eta $-pairing eigenstates in
the Hubbard model \cite{Yang1989}. The exact triplet-pair condensed
eigenstates of Kitaev model are also obtained in one and two-dimensional
systems \cite{MES_PRB1,MES_PRB_2}. As for interacting bosons, it has been
shown that a nearest-neighbour (NN) interaction may counteract the hardcore
effect, resulting in exact condensate eigenstates \cite%
{ZCH_PRB2025,zhang2025coalescing}. The similar behavior can never occur in a
Fermi system, although hardcore bosons and fermions in a chain share the
same spectrum. Notably, when only pair states are considered, both systems
may share similar eigenstates, since a local hardcore Bose pair and a Fermi
pair obey the same statistics. We schematically illustrate this point in
Fig. \ref{fig1}.

Nowadays, this topic is no longer merely conceptual. The related theoretical
study has been greatly advanced by cold-atom experiments. Advances in
cooling and trapping atoms and molecules with dipolar electric or magnetic
moments enable the realization of extended Hubbard models featuring
density-density interactions \cite%
{goral2002quantum,moses2015creation,moses2017new,baier2016extended,reichsollner2017quantum,Chomaz_2023}%
. Moreover, contemporary experimental setups enable precise control over
both the geometry and interactions, allowing for the direct investigation of
the real-time evolution of quantum many-body systems using engineered model
Hamiltonians \cite{jane2003simulation,bloch2012quantum,blatt2012quantum}. In
this scenario, a boson within the optical lattice essentially corresponds to
a cluster comprising an even number of fermions. This should lead to on-site
repulsive interactions within the framework of the tight-binding
description, causing an atom to become a hardcore boson in the strong
interaction limit.

In addition, an other related concept is the Hilbert space fragmentation
(HSF). It originates from intrinsic kinetic constraints \cite%
{schecter2019weak,yang2020hilbert,moudgalya2022hilbert,li2023hilbert,francica2023hilbert,nicolau2023flat}%
, which fragment the Hilbert space into dynamically isolated subspaces,
thereby rendering some states inaccessible and preventing full
thermalization. Constrained models, such as the PXP model \cite%
{lesanovsky2012interacting,turner2018weak}, constrained spin chains \cite%
{lingenfelter2024exact}, and dipole-conserving hopping models \cite%
{sala2020ergodicity}, were the first to exhibit fragmented dynamics, which
is indicative of HSF. In systems with fragmented Hilbert spaces, certain
subspaces may contain special eigenstates that are the quantum many-body
scars (QMBS) \cite%
{Shiraishi2017,Moudgalya2018,moudgalya20182,Khemani2019,Ho2019,Shibata2020,McClarty2020,Richter2022,Jeyaretnam2021,Turner2018,turner2018weak,Shiraishi2019,Lin2019,Choi2019,Khemani2020,Dooley2020,Dooley2021}
. These non-thermal states are typically embedded within the bulk spectrum
of the system and span a subspace in which initial states fail to thermalize
and instead exhibit periodic behavior.\textbf{\ }In practice, the kinetic
constraint is usually induced by particle--particle interactions rather than
by the intrinsic statistics of the bosons and fermions. Unlike the
exclusionary nature of the Pauli principle for fermions, the hardcore
constraint can lead to HSF.

In this work, we study both fermionic and bosonic extended Hubbard ladders.
To this end, we develop a general method to construct condensate eigenstates
from those of sub-Hamiltonians on plaquettes. A set of exact condensate-pair
eigenstates for the fermionic ladder is constructed by using the spectrum
generating algebra (SGA) \cite{Gadella2013}, thanks to the SU(2) symmetry of
the system. We then turn to the corresponding hardcore boson counterpart,
obtained simply by replacing fermionic operators with hardcore bosonic ones.
This system no longer possesses SU(2) symmetry; nevertheless, we find that
the boson-pair eigenstates can still be obtained because the model satisfies
the conditions of the restricted spectrum generating algebra (RSGA) \cite%
{moudgalya2020eta}. This follows directly from the fact that a local
hardcore Bose pair and a Fermi pair obey the same statistics. In addition,
we investigate the stability of both types of pair-condensate states against
perturbations introduced by next-nearest-neighbor (NNN) hopping, using
numerical simulations of the quench process. The conclusions can be extended
to a two-layer system. Our results on the one hand reveal the resemblance
between fermions and hardcore bosons upon pair formation; on the other hand,
they provide an accessible example of HSF in hardcore systems.

This paper is organized as follows. In Sec. \ref{General formalism}, we
present a theorem, based on which, the condensate eigenstates of a
Hamiltonian can be constructed from those of sub-Hamiltonians. The
sub-Hamiltonians are either possess a symmetry or meet the condition of
RSGA. In Secs. \ref{Fermi-Hubbard systems}, and \ref{Bose-Hubbard systems}\
we apply the theorem to fermi and hardcore-boson ladders, respectively. In
Sec. \ref{Dynamic stability under perturbation}, we conduct numerical
simulations to investigate the dynamic stability of the condensate states
under the NNN perturbations. Finally, we present a summary of our results in
Sec. \ref{Summary}.

\section{General formalism}

\label{General formalism}

Before proceeding to the concrete Hamiltonians, we present a general method
for obtaining exact eigenstates from those of small subsystems. This method
was first proposed in Ref. \cite{LSJY_PRB2025,Zhang_2025} and will be
further developed in the present work by establishing its connection to the
SGA and RSGA. This approach is applicable to more generalized fermion and
boson systems, with no specific restrictions on dimensionality or geometry.

We focus on many-body discrete quantum systems, including quantum spin,
interacting fermion and boson systems. Considering a Hamiltonian on a set of
lattice sites $(a,b,c)$, consisting two sub-Hamiltonians, given by $%
H(a,b,c)=H_{1}(a,c)+H_{2}(b,c)$, where $a$, $b$, and $c$ label three
sub-lattices with arbitrary sizes. The basic condition for the Hamiltonians
is that $H_{1}$ and $H_{2}$ possess a set of $\left( N+1\right) $-fold
degenerate zero-energy eigenstates in the ladder form, that is $%
H_{1}(s_{a}+s_{c})^{m}\left\vert G\right\rangle =0$ and $%
H_{2}(s_{b}+s_{c})^{m}\left\vert G\right\rangle =0$\ with $m\in \left[ 0,N%
\right] $, where $\left\vert G\right\rangle $\ is a common eigenstates of $%
H_{1}$ and $H_{2}$ on lattice sites $(a,b,c)$. Here, $s_{\alpha }$\ ($\alpha
=a,b,c$) is an operator on sub-lattice $\alpha $, which is an arbitrary
function of the spin operator and the particle creation and annihilation
operators within that sub-lattice. In practice, such Hamiltonians $H_{i}$
can be discovered and verified by exact diagonalization for small-size
system. For instance, the exact $\eta $-pairing eigenstates of a Hubbard
model are a paradigm to demonstrate this feature \cite{Yang1989}. Additional
examples are provided in recent works \cite%
{ma2022steady,ZCH_PRB2025,Ma_2024,MES_PRB1,MES_PRB_2,zhang2025coalescing,HDK_PRB2025}%
.

The underlying mechanism is twofold. We have the following conclusions,
respectively. (i) When the Hamiltonian terms commute with the combined
operators, 
\begin{equation}
\left[ H_{1},s_{a}+s_{c}\right] =0,\left[ H_{2},s_{b}+s_{c}\right] =0,
\end{equation}%
the SGA can be applied directly. We immediately have%
\begin{equation}
\left[ H,s_{a}+s_{b}+s_{c}\right] =0.
\end{equation}%
(ii) When the commutators are non-zero but annihilate the state $\left\vert
G\right\rangle $,%
\begin{equation}
\left[ H_{1},s_{a}+s_{c}\right] \left\vert G\right\rangle =\left[
H_{2},s_{b}+s_{c}\right] \left\vert G\right\rangle =0,
\end{equation}%
and the double commutators vanish,%
\begin{eqnarray}
\left[ \left[ H_{1},s_{a}+s_{c}\right] ,s_{a}+s_{c}\right] &=&0, \\
\left[ \left[ H_{2},s_{b}+s_{c}\right] ,s_{a}+s_{c}\right] &=&0,
\end{eqnarray}%
the RSGA is applicable. We immediately have%
\begin{equation}
\left[ H,s_{a}+s_{b}+s_{c}\right] \left\vert G\right\rangle =0,
\end{equation}%
and 
\begin{equation}
\left[ \left[ H,s_{a}+s_{b}+s_{c}\right] ,s_{a}+s_{b}+s_{c}\right] =0.
\end{equation}%
In both cases, there exists a set of zero-energy eigenstates of $H$
constructed via the operator $s_{a}+s_{b}+s_{c}$, given by%
\begin{equation}
H(a,b,c)(s_{a}+s_{b}+s_{c})^{m}\left\vert G\right\rangle =0.
\end{equation}%
We have the following remarks. First, the explicit form of the operators $%
\left\{ s_{a},s_{b},s_{c}\right\} $ is unrestricted, so our conclusion can
be generalized to high-dimensional systems. Second, it applies to
multi-sub-lattice systems. Third, although we only consider first-order
RSGA, higher-order RSGA is also applicable.

In this work we focus on three types of systems described by the following
general Hamiltonians. The first is a spinless Fermi--Hubbard model with
Hamiltonian

\begin{equation}
H_{\text{\textrm{F}}}=\sum_{i,j}\left( J_{ij}c_{i}^{\dag }c_{j}+\mathrm{H.c.}%
+V_{ij}n_{i}n_{j}\right) ,
\end{equation}%
where the fermionic operators satisfy%
\begin{equation}
\left\{ c_{l},c_{l^{\prime }}^{\dagger }\right\} =\delta _{l,l^{\prime
}},\left\{ c_{l},c_{l^{\prime }}\right\} =0.
\end{equation}%
Here $J_{ij}$ denotes the hopping amplitude for $i\neq j$ and the on-site
chemical potential for $i=j$, while $V_{ij}$ ($i\neq j$) gives the
density--density interaction strength. The second is its bosonic
counterpart, described by the Hamiltonian 
\begin{eqnarray}
H_{\text{\textrm{B}}} &=&\sum_{i,j}\left( J_{ij}b_{i}^{\dag }b_{j}+\mathrm{%
H.c.}+V_{ij}b_{i}^{\dag }b_{i}b_{j}^{\dag }b_{j}\right)   \notag \\
&&+\frac{1}{2}\sum_{j}U_{j}b_{j}^{\dag }b_{j}\left( b_{j}^{\dag
}b_{j}-1\right) ,
\end{eqnarray}%
where the bosonic operators satisfy%
\begin{equation}
\left[ b_{l},b_{l^{\prime }}^{\dagger }\right] =\delta _{l,l^{\prime }},%
\left[ b_{l},b_{l^{\prime }}\right] =0.
\end{equation}%
The additional term represents on-site interaction of strength $U_{j}$. The
third one is the Hamiltonian $H_{\text{\textrm{B}}}$\ in infinite $U_{j}$
limit, described by the Hamiltonian%
\begin{equation}
H_{\text{\textrm{HB}}}=\sum_{i,j}\left( J_{ij}a_{i}^{\dag }a_{j}+\mathrm{H.c.%
}+V_{ij}a_{i}^{\dag }a_{i}a_{j}^{\dag }a_{j}\right) ,
\end{equation}%
where the hardcore bosonic operators satisfy%
\begin{equation}
\left\{ a_{l},a_{l}^{\dagger }\right\} =1,\left\{ a_{l},a_{l}\right\} =0,
\end{equation}%
and%
\begin{equation}
\left[ a_{j},a_{l}^{\dagger }\right] =0,\left[ a_{j},a_{l}\right] =0,
\end{equation}%
for $j\neq l$.

Apparently $H_{\text{\textrm{F}}}$\ more closely resembles $H_{\text{\textrm{%
HB}}}$. According to our previous study \cite{ZCH_PRB2025}, although neither
hardcore bosons nor fermions can occupy the same single-site state, they
still obey different statistics, resulting in distinct many-particle quantum
states. Condensation is allowed for the former but forbidden for the latter
until Fermi pairs appear. Therefore, $H_{\text{\textrm{F}}}$ has no Fermi
counterparts to the condensate states found in $H_{\text{\textrm{HB}}}$.
However, the counterpart relation may emerge when only pair states are
considered. $H_{\text{\textrm{F}}}$ and $H_{\text{\textrm{HB}}}$ share the
same pair condensate, because a local hardcore Bose pair and a Fermi pair
obey identical statistics. In the following sections, we will demonstrate
this via exact solutions for concrete systems. Our strategy is to
investigate two parallel models, spinless Fermi and hardcore Bose Hubbard
models, which can be mapped onto each other by simply exchanging the
corresponding particle operators, i.e., $c_{j}\longleftrightarrow a_{j}$.

\section{Fermi-Hubbard systems}

\label{Fermi-Hubbard systems}

In this section we focus on the spinless Fermi--Hubbard model on a ladder
with interaction across the rung. We first obtain the exact solution for a
plaquette by direct diagonalization, then extend the results to large
systems using the method presented in the previous section.

\begin{figure}[th]
\centering
\includegraphics[width=0.48\textwidth]{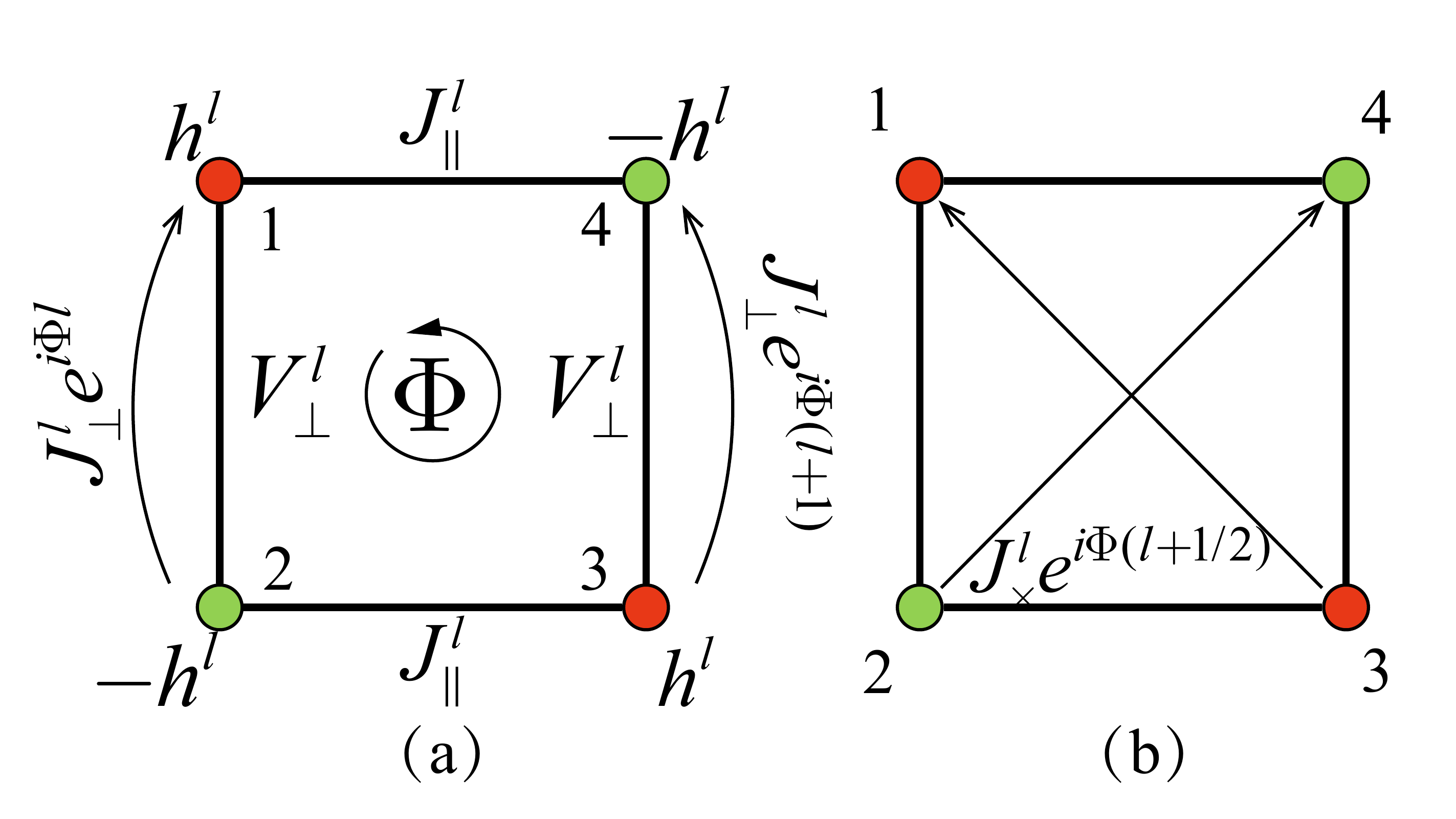}
\caption{Schematic of the lattice structures associated with the Fermi and
hardcore Bose Hamiltonians studied in this work. The plaquette corresponding
to the Hamiltonians $H_{\text{\textrm{F}}}^{l}$\ and $H_{\text{\textrm{B}}%
}^{l}$ given in Eqs. (\protect\ref{H_l,F}) and (\protect\ref{H_l,B}),
respectively. All parameters, including hopping amplitudes, nearest-neighbor
interaction strengths, and on-site potentials, are indicated in the panel.
The complex hopping amplitudes induce a magnetic flux threading the
plaquette.}
\label{fig2}
\end{figure}

\subsection{Fermi-Hubbard plaquette}

\label{Fermi-Hubbard plaquette}

We start with a Fermi--Hubbard model on a plaquette, with the Hamiltonian 
\begin{align}
H_{\text{\textrm{F}}}^{l}& =J_{\Vert }^{l}\left( c_{l,2}^{\dag
}c_{l,3}+c_{l,1}^{\dag }c_{l,4}\right)  \notag \\
& +J_{\perp }^{l}\left[ e^{i\Phi l}c_{l,1}^{\dag }c_{l,2}+e^{i\Phi \left(
l+1\right) }c_{l,4}^{\dag }c_{l,3}\right] +\mathrm{H.c.}  \notag \\
& +V_{\perp }^{l}\left[ \left( n_{l,1}-1/2\right) \left( n_{l,2}-1/2\right)
-1/4\right]  \notag \\
& +V_{\perp }^{l}\left[ \left( n_{l,3}-1/2\right) \left( n_{l,4}-1/2\right)
-1/4\right]  \notag \\
& +h^{l}\left( n_{l,1}-n_{l,2}+n_{l,3}-n_{l,4}\right) .  \label{H_l,F}
\end{align}%
Without loss of generality, we consider various terms including the hopping
between any two sites, the magnetic flux piercing the plaquette, the
density--density interaction, and the on-site potentials. The corresponding
parameters $\left\{ J_{\Vert }^{l},J_{\perp }^{l},\Phi ,V_{\perp
}^{l},h^{l}\right\} $ can be taken as any real number. The index $l$ is
already built into the Hamiltonian for later convenience.\ In Fig. \ref{fig2}%
(a), the Hamiltonian of the pluqette is schematically illustrated.

The Hamiltonian $H_{\text{\textrm{F}}}^{l}$ conserves particle number;
within each invariant subspace all eigenstates are obtained by exact
diagonalization. Nevertheless, we wish to obtain the exact solution of
interest in an alternative way.

Direct derivation shows that%
\begin{equation}
\left[ H_{\text{\textrm{F}}}^{l},\eta _{\text{\textrm{F}}}^{l}\right] =0,
\end{equation}%
where the operator is given by%
\begin{equation}
\eta _{\text{\textrm{F}}}^{l}=\left( c_{l,1}^{\dag }c_{l,2}^{\dag
}-c_{l,4}^{\dag }c_{l,3}^{\dag }\right) .
\end{equation}%
We find three degenerate eigenstates%
\begin{eqnarray}
\left\vert \phi _{n}^{l}\right\rangle &=&\frac{1}{n!\sqrt{C_{2}^{n}}}\left(
\eta _{\text{\textrm{F}}}^{l}\right) ^{n}\left\vert 0\right\rangle , \\
\left\vert \phi _{0}^{l}\right\rangle &=&\frac{1}{\sqrt{0!}}\left( \eta _{%
\text{\textrm{F}}}^{l}\right) ^{0}\left\vert 0\right\rangle , \\
\left\vert \phi _{1}^{l}\right\rangle &=&\frac{1}{1!\sqrt{C_{2}^{1}}}\left(
\eta _{\text{\textrm{F}}}^{l}\right) ^{1}\left\vert 0\right\rangle , \\
\left\vert \phi _{2}^{l}\right\rangle &=&\frac{1}{2!\sqrt{C_{2}^{2}}}\left(
\eta _{\text{\textrm{F}}}^{l}\right) ^{2}\left\vert 0\right\rangle ,
\end{eqnarray}%
all satisfying%
\begin{equation}
H_{\text{\textrm{F}}}^{l}\left\vert \phi _{0}^{l}\right\rangle =H_{\text{%
\textrm{F}}}^{l}\left\vert \phi _{1}^{l}\right\rangle =H_{\text{\textrm{F}}%
}^{l}\left\vert \phi _{2}^{l}\right\rangle =0,
\end{equation}%
where $\left\vert 0\right\rangle $\ is the vacuum state of the fermionic
operators. The eigenstates $\left\vert \phi _{j}^{l}\right\rangle $ ($%
j=0,1,2 $) are independent of the parameters $\left\{ J_{\Vert
}^{l},J_{\perp }^{l},\Phi ,V_{\perp }^{l},h^{l}\right\} $.

Now we extend the result to a $2\times 4$ plaquette with the Hamiltonian $H_{%
\text{\textrm{F}}}^{l}+H_{\text{\textrm{F}}}^{l+1}$ by setting $%
c_{l,4}=c_{l+1,1}$ and $c_{l,3}=c_{l+1,2}$. We apply the method presented in
the previous section, setting 
\begin{equation}
H_{1}=H_{\text{\textrm{F}}}^{l},H_{2}=H_{\text{\textrm{F}}}^{l+1}
\end{equation}%
and%
\begin{eqnarray}
s_{a} &=&c_{l,1}^{\dag }c_{l,2}^{\dag }, \\
s_{b} &=&c_{l+1,4}^{\dag }c_{l+1,3}^{\dag }, \\
s_{c} &=&-c_{l,4}^{\dag }c_{l,3}^{\dag }=-c_{l+1,1}^{\dag }c_{l+1,2}^{\dag }.
\end{eqnarray}%
Subsequently, we have%
\begin{equation}
\left[ H_{\text{\textrm{F}}}^{l}+H_{\text{\textrm{F}}}^{l+1},\eta _{\text{%
\textrm{F}}}^{l,l+1}\right] =0,
\end{equation}%
with 
\begin{eqnarray}
\eta _{\text{\textrm{F}}}^{l,l+1} &=&\left (c_{l,1}^{\dag }c_{l,2}^{\dag
}-c_{l,4}^{\dag }c_{l,3}^{\dag }+c_{l+1,4}^{\dag }c_{l+1,3}^{\dag }\right) 
\notag \\
&=&\left(c_{l,1}^{\dag }c_{l,2}^{\dag }-c_{l+1,1}^{\dag }c_{l+1,2}^{\dag
}+c_{l+1,4}^{\dag }c_{l+1,3}^{\dag }\right).
\end{eqnarray}%
Accordingly, we obtain a set of degenerate eigenstates%
\begin{equation}
\left\vert \phi _{n}^{l,l+1}\right\rangle =\frac{1}{n!\sqrt{C_{3}^{n}}}%
\left( \eta _{\text{\textrm{F}}}^{l,l+1}\right) ^{n}\left\vert
0\right\rangle ,
\end{equation}%
with $n=0,1,2,3$, satisfying%
\begin{equation}
\left( H_{\text{\textrm{F}}}^{l}+H_{\text{\textrm{F}}}^{l+1}\right)
\left\vert \phi _{n}^{l,l+1}\right\rangle =0.
\end{equation}%
This approach likewise allows the result to be extended to a ladder geometry
and, indeed, to certain bilayer systems. We will illustrate this point using
a ladder model.

\subsection{$\protect\eta $-pairing states in Fermi-Hubbard ladder}

\label{Fermi-Hubbard ladder}

We consider a ladder system with the Hamiltonian

\begin{eqnarray}
H_{\text{\textrm{FL}}} &=&\sum_{j}(J_{\Vert }\sum_{m=1,2}c_{j,m}^{\dag
}c_{j+1,m}+J_{\perp }\,e^{i\Phi j}c_{j,1}^{\dag }c_{j,2}  \notag \\
&&+\mathrm{H.c.}+V_{\perp }n_{j,1}n_{j,2}),  \label{H_FL,}
\end{eqnarray}%
which describes a uniform system threaded with a uniform magnetic flux \cite%
{Calvanese2017}.

\begin{figure}[th]
\centering
\includegraphics[width=0.48\textwidth]{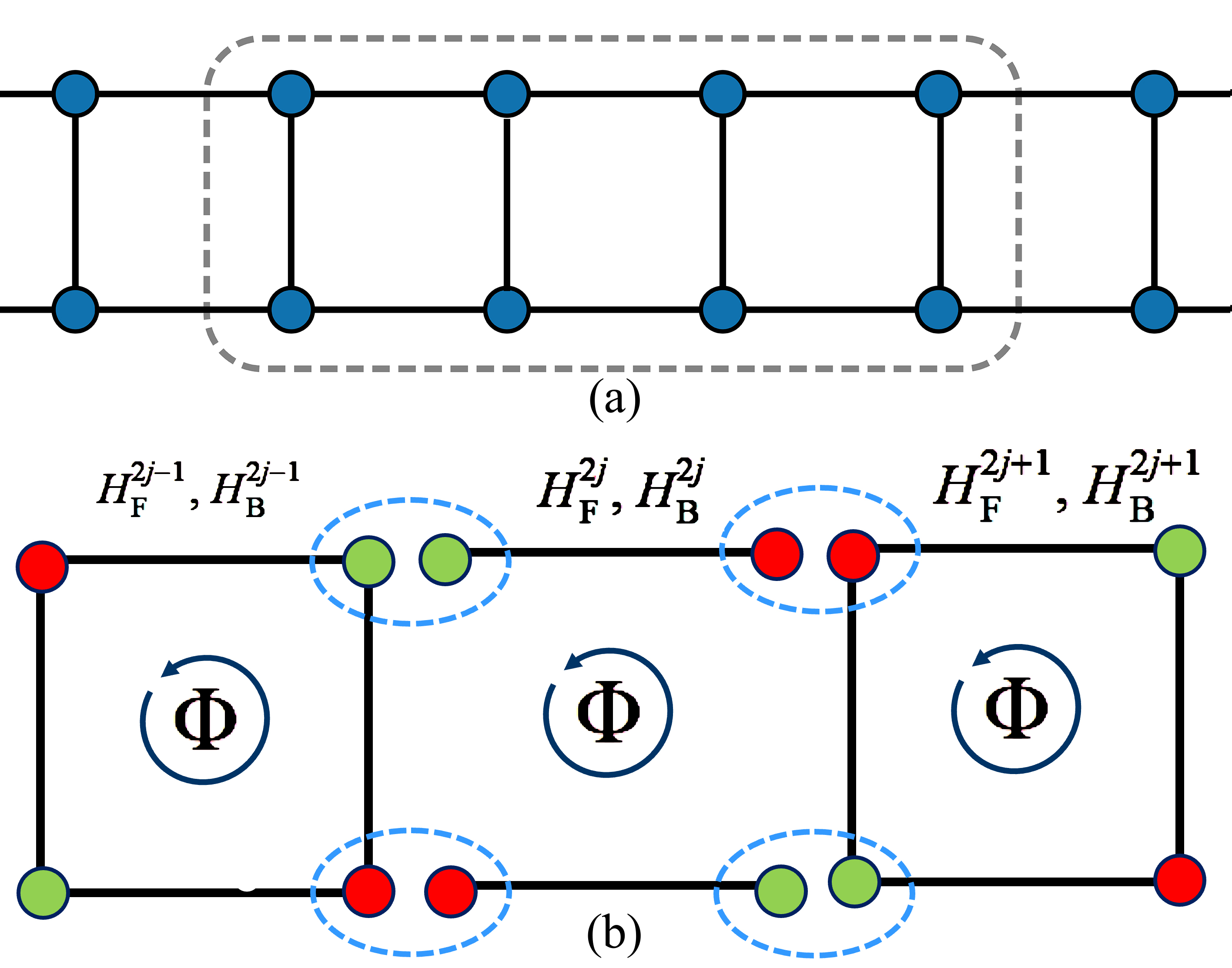}  
\caption{Schematic illustration of the decomposition of a ladder structure
associated with the Fermi and hardcore Bose Hamiltonians studied in this
work. (a) The ladder Hamiltonians $H_{\text{\textrm{FL}}}$\ and $H_{\text{ 
\textrm{BL}}}$ given in Eqs. (\protect\ref{H_FL,}) and (\protect\ref{H_HBL}
), respectively, can be expressed as the sum of two types of plaquettes
given in Eqs. (\protect\ref{H_FL_odd}, \protect\ref{H_FL_even}) and (\protect
\ref{H_HBL_odd}, \protect\ref{H_HBL_even}), respectively. (b) Schematic
illustration of the two types of plaquettes, the dotted circle encloses the
two combined sites.}
\label{Fig3}
\end{figure}

Considering the case with even $N$, $H_{\text{\textrm{FL}}}$\ can be written
as the sum of $N/2$ sub-Hamiltonians $H_{\text{\textrm{FL}}}^{l}$, given by%
\begin{equation}
H_{\text{\textrm{FL}}}=\sum_{l=1}^{N/2}H_{\text{\textrm{FL}}}^{l}.
\end{equation}%
Each sub-Hamiltonian describes a plaquette, and is explicitly expressed as 
\begin{align}
H_{\text{\textrm{FL}}}^{l}& =J_{\Vert }\sum_{m=1,2}c_{l,m}^{\dag
}c_{l+1,m}+J_{\perp }\,e^{i\Phi l}c_{l,1}^{\dag }c_{l,2}  \notag \\
& +J_{\perp }\,e^{i\Phi \left( l+1\right) }c_{l+1,1}^{\dag }c_{l+1,2}+%
\mathrm{H.c.}  \notag \\
& +V_{\perp }n_{l,1}n_{l,2}+V_{\perp }n_{l+1,1}n_{l+1,2},  \label{H_FL_odd}
\end{align}%
for odd $l$, while%
\begin{equation}
H_{\text{\textrm{FL}}}^{l}=J_{\Vert }\sum_{m=1,2}c_{l,m}^{\dag }c_{l+1,m}+%
\mathrm{H.c.},  \label{H_FL_even}
\end{equation}%
for even $l$. In Fig. \ref{Fig3}, this expression is schematically
illustrated.

Introducing the $\eta $-pairing operator 
\begin{equation}
\eta _{\text{\textrm{F}}}=\sum_{j=1}^{N}(-1)^{j}c_{j,1}^{\dag }c_{j,2}^{\dag
},
\end{equation}%
which satisfies the relation%
\begin{equation}
\left[ H_{\text{\textrm{FL}}},\eta _{\text{\textrm{F}}}\right] =0.
\end{equation}%
It indicates that the Hamiltonian possesses the SU(2) symmetry. We then can
construct the states%
\begin{equation}
\left\vert \psi _{n}^{\text{\textrm{F}}}\right\rangle =\frac{1}{n!\sqrt{%
C_{N}^{n}}}\left( \eta _{\text{\textrm{F}}}\right) ^{n}\left\vert
0\right\rangle ,
\end{equation}%
using SGA. Based on the fact that%
\begin{equation}
\left\langle \psi _{m}^{\text{\textrm{F}}}\right\vert c_{j,1}^{\dag
}c_{j,2}^{\dag }\left\vert \psi _{n}^{\text{\textrm{F}}}\right\rangle
=\left( -1\right) ^{j}\frac{\sqrt{\left( N-n\right) \left( n+1\right) }}{N}%
\delta _{m-1,n}
\end{equation}%
we obtain the correlation function as%
\begin{equation}
\left\langle \psi _{n}^{\text{\textrm{F}}}\right\vert c_{j,1}^{\dag
}c_{j,2}^{\dag }c_{j+r,1}c_{j+r,2}\left\vert \psi _{n}^{\text{\textrm{F}}%
}\right\rangle =\left( -1\right) ^{r}\frac{\left( N-n\right) n}{N(N-1)}.
\label{<cccc>}
\end{equation}%
This shows that the state $\left\vert \psi _{n}^{\text{\textrm{F}}%
}\right\rangle $ possesses off-diagonal long range order (ODLRO), as shown
in Ref. \cite{yang1962concept}, because the correlation function does not
decay with increasing $r$. To understand the features of the state $%
\left\vert \psi _{n}^{\text{\textrm{F}}}\right\rangle $, we consider its
similarity to the Bose-condensate state with $\pi $ momentum,%
\begin{equation}
\left\vert \chi _{n}\right\rangle =\frac{1}{\sqrt{n!}}\left( \frac{1}{\sqrt{N%
}}\sum_{j=1}^{N}e^{i\pi j}b_{j}^{\dag }\right) ^{n}\left\vert 0\right\rangle
.
\end{equation}%
The corresponding correlation function is%
\begin{equation}
\left\langle \chi _{n}\right\vert b_{l}^{\dag }b_{l+r}\left\vert \chi
_{n}\right\rangle =e^{-i\pi r}\frac{n}{N},  \label{<bb>}
\end{equation}%
indicating ODLRO. Comparing this with Eq. (\ref{<cccc>}), we see that the
two correlators coincide in the dilute limit $n/N\ll 1$ and remain
qualitatively alike at finite density. Thus $\left\vert \psi _{n}^\text{%
\textrm{F}}\right\rangle $ and $\left\vert \chi _{n}\right\rangle $ exhibit
the same condensate character.

\section{Bose-Hubbard systems}

\label{Bose-Hubbard systems}

In this section we investigate the hardcore Bose-Hubbard model, the
counterpart of the spinless Fermi system. In parallel, we first examine a
hardcore Bose-Hubbard plaquette and construct the pairing eigenstates by
using RSGA. We then turn to the hardcore Bose-Hubbard ladder.

\subsection{Hardcore Bose-Hubbard plaquette}

\label{Hubbard plaquette}

We start with a hardcore Bose--Hubbard plaquette whose Hamiltonian is 
\begin{align}
H_{\text{\textrm{B}}}^{l}& =J_{\Vert }^{l}\left( a_{l,2}^{\dag
}a_{l,3}+a_{l,1}^{\dag }a_{l,4}\right)  \notag \\
& +J_{\perp }^{l}\left[ e^{i\Phi l}a_{l,1}^{\dag }a_{l,2}+e^{i\Phi \left(
l+1\right) }a_{l,4}^{\dag }a_{l,3}\right] +\mathrm{H.c.}  \notag \\
& +V_{\perp }^{l}\left[ \left( a_{l,1}^{\dag }a_{l,1}-1/2\right) \left(
a_{l,2}^{\dag }a_{l,2}-1/2\right) -1/4\right]  \notag \\
& +V_{\perp }^{l}\left[ \left( a_{l,3}^{\dag }a_{l,3}-1/2\right) \left(
a_{l,4}^{\dag }a_{l,4}-1/2\right) -1/4\right]  \notag \\
& +h^{l}\left( a_{l,1}^{\dag }a_{l,1}-a_{l,2}^{\dag }a_{l,2}+a_{l,3}^{\dag
}a_{l,3}-a_{l,4}^{\dag }a_{l,4}\right) .  \label{H_l,B}
\end{align}%
This expression is obtained from $H_{\text{\textrm{F}}}^{l}$\ by replacing $%
c_{j}$ with $a_{j}$.

Direct derivation shows that although the commutators $\left[ H_{\text{%
\textrm{B}}}^{l},\eta _{\text{\textrm{B}}}^{l}\right] $\ iself is non-zero,
it annihilates the vacuum state of the bosonic operators.%
\begin{equation}
\left[ H_{\text{\textrm{B}}}^{l},\eta _{\text{\textrm{B}}}^{l}\right]
\left\vert \text{Vac}\right\rangle =0,
\end{equation}%
and the double commutator vanishes%
\begin{equation}
\left[ \left[ H_{\text{\textrm{B}}}^{l},\eta _{\text{\textrm{B}}}^{l}\right]
,\eta _{\text{\textrm{B}}}^{l}\right] =0,
\end{equation}%
\ with%
\begin{equation}
\eta _{\text{\textrm{B}}}^{l}=\left( a_{l,1}^{\dag }a_{l,2}^{\dag
}-a_{l,4}^{\dag }a_{l,3}^{\dag }\right) .
\end{equation}%
These conditions guarantee that RSGA applies, yielding three degenerate
eigenstates%
\begin{eqnarray}
\left\vert \varphi _{n}^{l}\right\rangle &=&\frac{1}{n!\sqrt{C_{2}^{n}}}%
\left( \eta _{\text{\textrm{B}}}^{l}\right) ^{n}\left\vert \text{Vac}%
\right\rangle , \\
\left\vert \varphi _{0}^{l}\right\rangle &=&\frac{1}{\sqrt{0!}}\left( \eta _{%
\text{\textrm{B}}}^{l}\right) ^{0}\left\vert \text{Vac}\right\rangle , \\
\left\vert \varphi _{1}^{l}\right\rangle &=&\frac{1}{1!\sqrt{C_{2}^{1}}}%
\left( \eta _{\text{\textrm{B}}}^{l}\right) ^{1}\left\vert \text{Vac}%
\right\rangle , \\
\left\vert \varphi _{2}^{l}\right\rangle &=&\frac{1}{2!\sqrt{C_{2}^{2}}}%
\left( \eta _{\text{\textrm{B}}}^{l}\right) ^{2}\left\vert \text{Vac}%
\right\rangle ,
\end{eqnarray}%
all satisfying%
\begin{equation}
H_{\text{\textrm{B}}}^{l}\left\vert \varphi _{j}^{l}\right\rangle =0,\text{ }%
j=0,1,2.
\end{equation}%
The set of eigenstates $\left\vert \varphi _{j}^{l}\right\rangle $ is still
independent of the parameters $\left\{J_{\Vert }^{l},J_{\perp }^{l}, \Phi ,
V_{\perp }^{l},h^{l} \right\}$ .

Similarly, using the operator%
\begin{eqnarray}
\eta _{\text{\textrm{B}}}^{l,l+1} &=&\left( a_{l,1}^{\dag }a_{l,2}^{\dag
}-a_{l,4}^{\dag }a_{l,3}^{\dag }+a_{l+1,4}^{\dag }a_{l+1,3}^{\dag }\right) 
\notag \\
&=&\left( a_{l,1}^{\dag }a_{l,2}^{\dag }-a_{l+1,1}^{\dag }a_{l+1,2}^{\dag
}+a_{l+1,4}^{\dag }a_{l+1,3}^{\dag }\right) ,
\end{eqnarray}%
we construct the eigenstates of $H_{\text{\textrm{F}}}^{l}+H_{\text{\textrm{F%
}}}^{l+1}$, 
\begin{equation}
\left\vert \varphi _{n}^{l,l+1}\right\rangle =\frac{1}{n!\sqrt{C_{3}^{n}}}%
\left( \eta _{\text{\textrm{B}}}^{l,l+1}\right) ^{n}\left\vert
0\right\rangle ,
\end{equation}%
for $n=0,1,2,3$, all satisfying%
\begin{equation}
\left( H_{\text{\textrm{B}}}^{l}+H_{\text{\textrm{B}}}^{l+1}\right)
\left\vert \varphi _{n}^{l,l+1}\right\rangle =0.
\end{equation}%
As we have done for the Fermi system, this approach likewise allows the
result to be extended to a ladder geometry, and indeed, to certain bilayer
systems. We will illustrate this point using a ladder model.

\subsection{$\protect\eta $-pairing states in Hardcore Bose-Hubbard ladder}

\label{Hardcore Bose-Hubbard ladder}

\begin{figure*}[tbh]
\centering
\includegraphics[width=1\textwidth]{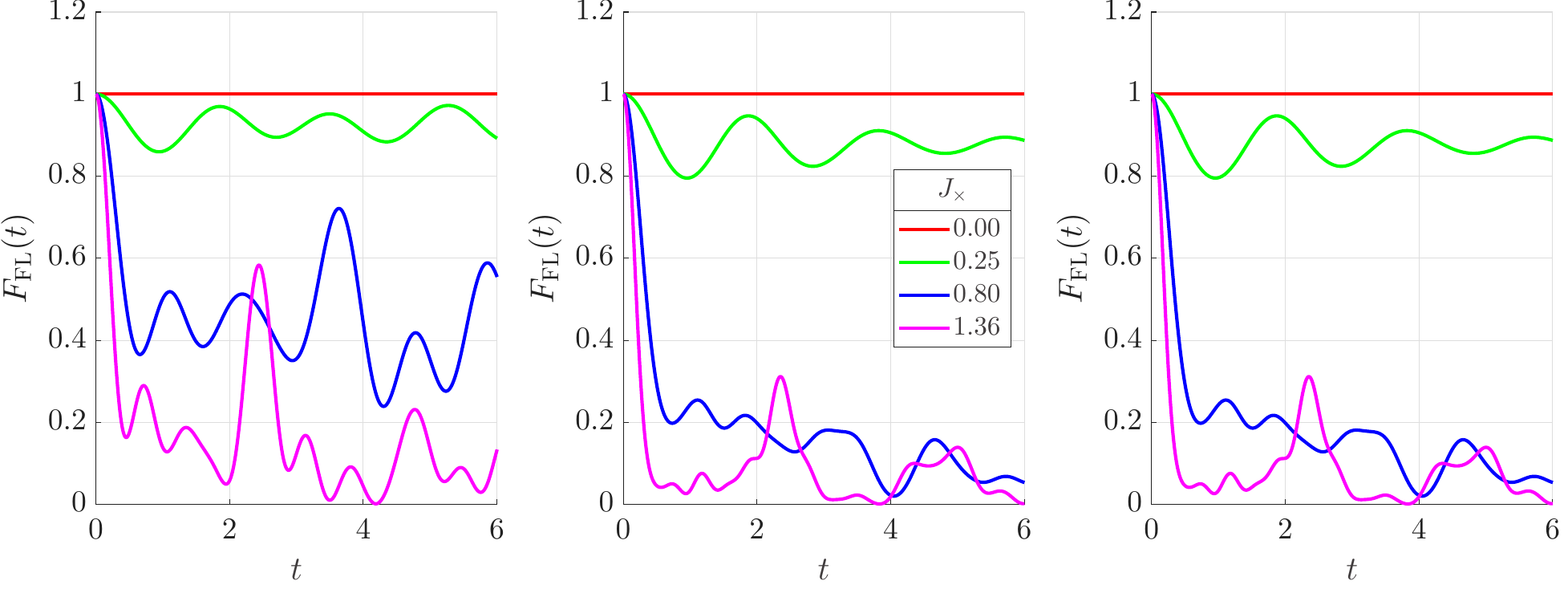}
\caption{ Fidelity dynamics of $\protect\eta $-pairing states in the
Fermi-Hubbard ladder under quench by next-nearest-neighbor hopping. The
initial states are the exact eigenstates $\left\vert \psi _{n}^{\text{\textrm{F}}%
}\right\rangle =(\protect\eta _{\mathrm{F}})^{n}|0\rangle /(n!\protect\sqrt{%
C_{N}^{n}})$ with $\protect\eta _{\mathrm{F}}=\sum_{j}(-1)^{j}c_{j,1}^{%
\dagger }c_{j,2}^{\dagger }$. Panels (a), (b), and (c) correspond to pair
numbers $n=1$, $n=2$, and $n=3$, respectively, for a system of $N=20$ rungs.
In each panel, different curves represent different strengths of the
next-nearest-neighbor hopping $J_{\times }$ added to the Hamiltonian in Eq. (%
\protect\ref{H_FL,Jx}), which explicitly breaks the SU(2) symmetry and the
RSGA condition. The time evolution is governed by $\mathcal{H}_{\mathrm{FL}%
}=H_{\mathrm{FL}}+J_{\times }\sum_{l}(c_{l,1}^{\dagger }c_{l+1,2}+\mathrm{%
H.c.})$, with fixed parameters $J_{\parallel }=1.32$, $J_{\perp }=0.64$, $%
V_{\perp }=3.14$ and $\Phi =0$. The fidelity $F_{\mathrm{FL}}(t)=|\langle 
\protect\psi _{n}^{\mathrm{F}}|e^{-i\mathcal{H}_{\mathrm{FL}}t}|\protect\psi %
_{n}^{\mathrm{F}}\rangle |^{2}$ decays rapidly for all $J\times \neq 0$,
confirming that $\left\vert \psi _{n}^{\text{\textrm{F}}%
}\right\rangle $ are no longer
eigenstates of the perturbed Hamiltonian. The decay rate increases with $%
J_{\times }$ and exhibits a weak dependence on the pair number $n$. }
\label{fig4}
\end{figure*}

In this section, we investigate a hardcore Bose-Hubbard ladder, which is the
counterpart of the spinless Fermi ladder. The corresponding Hamiltonian of a
hardcore Bose-Hubbard ladder is%
\begin{align}
H_{\text{\textrm{HBL}}}& =\sum_{j=1}^{N}(J_{\Vert }\sum_{m=1,2}a_{j,m}^{\dag
}a_{j+1,m}+J_{\perp }\,e^{i\Phi j}a_{j,1}^{\dag }a_{j,2}  \notag \\
& +\mathrm{H.c.}+V_{\perp }a_{j,1}^{\dag }a_{j,1}a_{j,2}^{\dag }a_{j,2}),
\label{H_HBL}
\end{align}%
with the hardcore constraint $\left( a_{j,m}\right) ^{2}=0$. Similarly,
considering the case with even $N$, $H_{\text{\textrm{HBL}}}$\ can be
written as the sum of $N/2$ sub-Hamiltonians $H_{\text{\textrm{HBL}}}^{l}$,
given by%
\begin{equation}
H_{\text{\textrm{HBL}}}=\sum_{l=1}^{N/2}H_{\text{\textrm{HBL}}}^{l}.
\end{equation}%
Each sub-Hamiltonian describes a plaquette, and is explicitly expressed as 
\begin{align}
H_{\text{\textrm{HBL}}}^{l}& =J_{\Vert }\sum_{m=1,2}a_{l,m}^{\dag
}a_{l+1,m}+J_{\perp }\,e^{i\Phi l}a_{l,1}^{\dag }a_{l,2}  \notag \\
& +J_{\perp }\,e^{i\Phi \left( l+1\right) }a_{l+1,1}^{\dag }a_{l+1,2}+%
\mathrm{H.c.}  \notag \\
& +V_{\perp }a_{l,1}^{\dag }a_{l,1}a_{l,2}^{\dag }a_{l,2}+V_{\perp
}a_{l+1,1}^{\dag }a_{l+1,1}a_{l+1,2}^{\dag }a_{l+1,2},  \label{H_HBL_odd}
\end{align}%
for odd $l$, while%
\begin{equation}
H_{\text{\textrm{HBL}}}^{l}=J_{\Vert }\sum_{m=1,2}a_{l,m}^{\dag }a_{l+1,m}+%
\mathrm{H.c.},  \label{H_HBL_even}
\end{equation}%
for even $l$.

Introducing the $\eta $-pairing operator 
\begin{equation}
\eta _{\text{\textrm{HB}}}=\sum_{j=1}^{N}\left( -1\right) ^{j}a_{j,1}^{\dag
}a_{j,2}^{\dag },
\end{equation}%
and applying the method presented in the previous section, we obtain the
eigenstates%
\begin{equation}
\left\vert \psi _{n}^{\text{\textrm{HB}}}\right\rangle =\frac{1}{n!\sqrt{%
C_{N}^{n}}}\left( \eta _{\text{\textrm{HB}}}\right) ^{n}\left\vert
0\right\rangle ,
\end{equation}%
which have the same form of $\left\vert \psi _{n}^{\text{\textrm{F}}%
}\right\rangle $. The correlation function is also identical to those for
Fermi system, i.e.,%
\begin{equation}
\left\langle \psi _{n}^{\text{\textrm{HB}}}\right\vert a_{j,1}^{\dag
}a_{j,2}^{\dag }a_{j+r,1}a_{j+r,2}\left\vert \psi _{n}^{\text{\textrm{HB}}%
}\right\rangle =\left( -1\right) ^{r}\frac{\left( N-n\right) n}{N(N-1)},
\end{equation}%
indicating that the state $\left\vert \psi _{n}^{\text{\textrm{HB}}%
}\right\rangle $ possesses ODLRO.

When $V_{\perp }$ equals zero, the eigenstate $\left\vert \psi _{n}^{\text{%
\textrm{HB}}}\right\rangle $ still exists. Note that there are many other
condensate eigenstates in the dilute limit $n/N\ll 1$, which can be obtained
from the single-particle eigenstates of the Hamiltonian. However, compared
to these single-particle condensate states, $\left\vert \psi _{n}^{\text{%
\textrm{HB}}}\right\rangle $ describes condensation of boson pairs.

\section{Dynamic stability under perturbation}

\label{Dynamic stability under perturbation}

It has been shown that the conclusion of the previous section holds under
either SGA or RSGA. Both models, however, retain only NN hopping and thus
assume NNN hopping to vanish. This assumption is rarely satisfied in real
materials, where the wave-function overlap between next-nearest neighbors is
almost always finite. Here we examine how finite NNN hopping influences the
existence of condensate states in both systems. Our strategy is first to
analyze how non-zero NNN hopping modifies the SGA and RSGA conditions for
Fermi and Bose plaquettes, and then to study the dynamical response of the
corresponding ladder condensates upon a quench.

We begin by adding NNN hopping terms to both Fermi and Bose plaquettes.

(i) For the Fermi plaquette, the resulting Hamiltonian is 
\begin{equation}
\mathcal{H}_{\text{\textrm{F}}}^{l}=H_{\text{\textrm{F}}}^{l}+J_{\times }^{l}%
\left[ e^{i\Phi \left( l+1/2\right) }\left( c_{l,1}^{\dag
}c_{l,3}+c_{l,4}^{\dag }c_{l,2}\right) +\mathrm{H.c.}\right] ,
\end{equation}%
where the NNN links are sketched in Fig. \ref{fig2}(b). Straightforward
calculation gives%
\begin{eqnarray}
\left[ \mathcal{H}_{\text{\textrm{F}}}^{l},\eta _{\text{\textrm{F}}}^{l}%
\right] &=&2J_{\times }^{l}\left[ e^{i\Phi \left( l+1/2\right)
}c_{l,1}^{\dag }c_{l,4}^{\dag }-e^{-i\Phi \left( l+1/2\right) }c_{l,2}^{\dag
}c_{l,3}^{\dag }\right]  \notag \\
&\neq &0, \\
\left[ \mathcal{H}_{\text{\textrm{F}}}^{l},\eta _{\text{\textrm{F}}}^{l}%
\right] \left\vert 0\right\rangle &\neq &0,
\end{eqnarray}%
and%
\begin{equation}
\left[ \left[ \mathcal{H}_{\text{\textrm{F}}}^{l},\eta _{\text{\textrm{F}}%
}^{l}\right] ,\eta _{\text{\textrm{F}}}^{l}\right] =0,
\end{equation}%
demonstrating that $\mathcal{H}_{\text{\textrm{F}}}^{l}$ no longer satisfies
either SGA (breaking the SU(2) symmetry) or RSGA.

(ii) For the Bose plaquette, the analogous Hamiltonian reads 
\begin{equation}
\mathcal{H}_{\text{\textrm{B}}}^{l}=H_{\text{\textrm{B}}}^{l}+J_{\times
}^{l} \left[ e^{i\Phi (l+1/2)}\left(a_{l,1}^{\dag }a_{l,3}+a_{l,4}^{\dag
}a_{l,2}\right) +\mathrm{H.c.}\right] .
\end{equation}%
Straightforward calculation gives%
\begin{equation}
\left[ \mathcal{H}_{\text{\textrm{B}}}^{l},\eta _{\text{\textrm{B}}}^{l}%
\right] \left\vert \text{Vac}\right\rangle =0,
\end{equation}%
and%
\begin{equation}
\left[ \left[ \mathcal{H}_{\text{\textrm{B}}}^{l},\eta _{\text{\textrm{B}}%
}^{l}\right] ,\eta _{\text{\textrm{B}}}^{l}\right] =0,
\end{equation}%
indicating that $\mathcal{H}_{\text{\textrm{B}}}^{l}$ still satisfies the
conditions of both SGA and RSGA. These results can be applied to Fermi and
Bose ladders upon adding NNN hopping terms. The resulting Fermi-ladder
Hamiltonian is 
\begin{eqnarray}
\mathcal{H}_{\text{\textrm{FL}}} &=&J_{\times }\sum_{l=1}^{N/2}\left[
e^{i\Phi \left( l+1/2\right) }\left( c_{l,1}^{\dag
}c_{l+1,2}+c_{l+1,1}^{\dag }c_{l,2}\right) +\mathrm{H.c.}\right]  \notag \\
&&+H_{\text{\textrm{FL}}}.  \label{H_FL,Jx}
\end{eqnarray}%
Because this Hamiltonian no longer satisfies the RSGA condition, the state $%
\left\vert \psi _{n}^{\text{\textrm{F}}}\right\rangle $ is not one of its
eigenstates.

Likewise, the corresponding Bose-ladder Hamiltonian becomes 
\begin{eqnarray}
\mathcal{H}_{\text{\textrm{HBL}}} &=&J_{\times }\sum_{l=1}^{N/2}\left[
e^{i\Phi \left( l+1/2\right) }\left( a_{l,1}^{\dag
}a_{l+1,2}+a_{l+1,1}^{\dag }a_{l,2}\right) +\mathrm{H.c.}\right]  \notag \\
&&+H_{\text{\textrm{HBL}}},
\end{eqnarray}%
which can be shown to still satisfy the RSGA condition. Consequently, the
state $\left\vert \psi _{n}^{\text{\textrm{HB}}}\right\rangle $ is immune to
the NNN hopping terms and remains an eigenstate of $\mathcal{H}_{\text{%
\textrm{HBL}}}$.

To demonstrate these results, we examine the dynamic response of the states $%
\left\vert \psi _{n}^{\text{\textrm{F}}}\right\rangle $\ and $\left\vert
\psi _{n}^{\text{\textrm{HB}}}\right\rangle $\ under a quantum quench.
Specifically, we compute their time evolution under the quench Hamiltonians $%
\mathcal{H}_{\text{\textrm{FL}}}$ and\ $\mathcal{H}_{\text{\textrm{HBL}}}$,
respectively. To quantify the response, we use the fidelities%
\begin{equation}
F_{\text{\textrm{FL}}}(t)=\left\vert \left\langle \psi _{n}^{\text{\textrm{F}%
}}\right\vert e^{-i\mathcal{H}_{\text{\textrm{FL}}}t}\left\vert \psi _{n}^{%
\text{\textrm{F}}}\right\rangle \right\vert ^{2},
\end{equation}%
and%
\begin{equation}
F_{\text{\textrm{HBL}}}(t)=\left\vert \left\langle \psi _{n}^{\text{\textrm{%
HB}}}\right\vert e^{-i\mathcal{H}_{\text{\textrm{HBL}}}t}\left\vert \psi
_{n}^{\text{\textrm{HB}}}\right\rangle \right\vert ^{2},
\end{equation}%
which characterize how the system departs from its initial state as a
function of the NNN hopping strength $J_{\times }$. Both quantities can be
obtained via exact diagonalization for finite systems. As shown above, we
have $F_{\text{\textrm{HBL}}}(t)=1$\ for all times. {We plot }$F_{\text{%
\textrm{FL}}}(t)${\ in Fig. \ref{fig4} as a function of }${t}${\ for
selected system sizess and particle numbers}. The results accord with our
predictions. The $F_{\text{\textrm{HBL}}}(t)$\ remains unchanged, while $F_{%
\text{\textrm{FL}}}(t)$ decays with time $t$ for nonzero values of $%
J_{\times }$ as expected.

\section{Summary and discussion}

\label{Summary}

In summary, we have extended the general method to construct the condensate
eigenstates for systems without symmetry but meeting the RGSA. This may pave
a way to find systems possessing HSF. We exemplified this finding through
the investigation of both fermionic and bosonic extended Hubbard ladders. A
set of exact condensate-pair eigenstates for the fermionic ladder is
constructed by using the SGA, and boson-pair eigenstates based on the RSGA.
This demonstrates one of the main points of this work: that although
fermions and hardcore bosons obey different statistics, a local hardcore
Bose pair and a Fermi pair obey the same statistics. In addition, we
investigate the stability of both types of pair-condensate states against
perturbations introduced by NNN hopping. We found that the NNN hopping terms
break the symmetry as well as the condition of RSGA, while the boson-pair
eigenstates remain unchanged. These are demonstrated by using numerical
simulations of the quench process. The conclusions can be extended to a
two-layer system. Furthermore, our results have important implications for
future investigations. The hardcore constraint plays a crucial role in the
existence of the set of eigenstates $\left\{ \left\vert \psi _{n}^{\text{%
\textrm{HB}}}\right\rangle \right\} $. Notably, when we release such an $N$%
-site $n$-boson-pair condensate in a large-sized ladder, the correlation of
pairs in the bulk of the condensate can enhance the stability of the pairing
configuration despite the collapse that may occur at the edges.
Consequently, this may lead to the HSF, since there is limited access to
other Hilbert space configurations. This provides an accessible example of
the HSF in hardcore systems, as guidance for experimental detection.

\section*{ACKNOWLEDGMENTS}

This paper was supported by the National Natural Science Foundation of China
(under Grant No. 12374461).

\bibliography{reference}

\end{document}